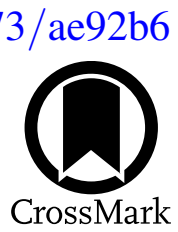

# First Very Long Baseline Interferometry Fringe Detection at 690 GHz

Ming-Tang Chen[1,2], Geoffrey B. Crew[3], Dan Bintley[4], Alan Roy[5], Keiichi Asada[1], Johnson Han[1], Kuan-Yu Liu[4], Chin-Shin Chang[6,7], Andrea McClöskey[4], Yiqing Song[6], Vincent Fish[3], Michael Lindqvist[8], Satoki Matsushita[1], Lynn D. Matthews[3], Hugo Messias[6], Jan F. Wagner[5], Ue-Li Pen[1], Derek Kubo[1,2], Teddy Huang[1], Philippe Raffin[1], Chih-Wei L. Huang[1], Pierre Martin-Cocher[1], Geoffrey C. Bower[1,4], Patrick Koch[1], Lisa Kewley[9], Timothy Norton[9], Nimesh Patel[9], Paul T. P. Ho[4], Per Friberg[4], Izumi Mizuno[4], Akil Marshall[4], Andrey M. Baryshev[10], Heino Falcke[11], Thomas Greve[12], Jacques Lepine[13,30], Manuel Fernández López[14,30], Silvina Cichowolski[15,30], Carlos Valotto[16,17,30], Ciriaco Goddi[18,19,20,30], Guillermo Giménez de Castro[21,22,30], Fátima Salete Correra[23,30], Marianne Vestergaard[24], Jes Kristian Jørgensen[25], Martin E. Pessah[26], and Roman Gold[27,28,29]

[1] Academia Sinica Institute of Astronomy & Astrophysics, Taipei 106, Taiwan; mtchen@asiaa.sinica.edu.tw
[2] Academia Sinica Institute of Astronomy & Astrophysics, Hawaii Operations, N. A'Ohoku Place, Hilo, HI 96720, USA
[3] Massachusetts Institute of Technology, Haystack Observatory, 99 Millstone Road, Westford, MA 01886, USA
[4] East Asian Observatory, 660 N. A'ohoku Pl., Hilo, HI 96720, USA
[5] Max-Planck-Institut für Radioastronomie, Auf dem Hügel 69, D-53121 Bonn, Germany
[6] Joint ALMA Observatory, Alonso de Córdova 3107, Vitacura 763-0355, Santiago, Chile
[7] Department of Astronomy, University of Geneva, Chemin Pegasi 51, 1290 Versoix, Switzerland
[8] Department of Physics and Astronomy, Chalmers University of Technology, Onsala Space Observatory, SE-43992 Onsala, Sweden
[9] Center for Astrophysics | Harvard & Smithsonian, 60 Garden St., Cambridge, MA 02138, USA
[10] NOVA Sub-mm Instrumentation Group, Kapteyn Astronomical Institute, University of Groningen, Landleven 12, 9747 AD Groningen, The Netherlands
[11] Department of Astrophysics, (IMAPP), Radboud University, PO Box 9010, 6500 GL Nijmegen, The Netherlands
[12] DTU Space, Technical University of Denmark. Building 328, Elektrovej, DK-2800 Kgs. Lyngby, Denmark
[13] Instituto de Astronomia, Geofísica e Ciências Atmosféricas, Universidade de São Paulo, São Paulo, SP 05508-090, Brazil
[14] Instituto Argentino de Radioastronomía, (CCT La Plata, CONICET; CICPBA; UNLP), C.C.5, (1894) Villa Elisa, Buenos Aires, Argentina
[15] Instituto de Astronomía y Física del Espacio, CONICET–UBA, Argentina
[16] Instituto de Astronomía Teórica y Experimental, CONICET-UNC, Córdoba, Argentina
[17] Observatorio Astronómico de Córdoba, Universidad Nacional de Córdoba, Laprida 854, Córdoba X5000BGR, Argentina
[18] INAF—Osservatorio Astronomico di Cagliari, via della Scienza 5, I-09047 Selargius (CA), Italy
[19] INFN, sezione di Cagliari, I-09042 Monserrato (CA), Italy
[20] Dipartimento di Fisica, Universitá degli Studi di Cagliari, SP Monserrato-Sestu km 0.7, I-09042 Monserrato (CA), Italy
[21] Universidade Presbiteriana Mackenzie, Centro de Rádio Astronomia e Astrofísica Mackenzie, São Paulo, Brazil
[22] Instituto de Astronomía y Física del Espacio, CONICET-UBA, Buenos Aires, Argentina
[23] Escola Politécnica da Universidade de São Paulo, Departamento de Engenharia de Sistemas Eletrônicos, Av. Prof. Luciano Gualberto, trv. 3, n. 158, São Paulo, SP, Brazil
[24] DARK, Niels Bohr Institute, University of Copenhagen, Denmark
[25] Niels Bohr Institute, University of Copenhagen, Jagtvej 155A, DK 2200 Copenhagen N., Denmark
[26] Niels Bohr International Academy, Niels Bohr Institute, Blegdamsvej 17, DK-2100 Copenhagen Ø, Denmark
[27] CP3-Origins, University of Southern Denmark, Campusvej 55, DK-5230 Odense, Denmark
[28] Institute for Mathematics and Interdisciplinary Center for Scientific Computing, Heidelberg University, Im Neuenheimer Feld 205, Heidelberg 69120, Germany
[29] Institut für Theoretische Physik, Universität Heidelberg, Philosophenweg 16, 69120 Heidelberg, Germany

## Abstract

We report the first very long baseline interferometry (VLBI) experiment conducted in the 690 GHz atmospheric window. On 2024 November 21, observations with the Atacama Large Millimeter/submillimeter Array, the Atacama Pathfinder EXperiment (APEX), and the James Clerk Maxwell Telescope (JCMT) were carried out using ALMA's newly developed Band 9 phasing capability. Fringes were detected on the ALMA–APEX baseline during a scan of the quasar J0423–0120, with a signal-to-noise ratio of ∼12 and useful fringe recovery over solution intervals of order tens of seconds, representing the highest-frequency ground-based VLBI fringe detection reported to date. No fringes were found on the ALMA–JCMT baseline, despite excellent weather conditions, consistent with sensitivity predictions and supporting baseline performance models. The ALMA

[30] The Large Latin American Millimeter Array collaboration.

Phasing System maintained stable phasing at Band 9 for ∼1–2 minutes before gradually degrading, indicating limitations under these observing conditions. Our analysis shows that, under excellent weather conditions, 690 GHz VLBI can still support fringe recovery over short solution intervals, despite rapid atmospheric phase fluctuations at these frequencies. This work validated key elements of near-terahertz VLBI operation and establishes a technical foundation for routine observations in the 690 GHz atmospheric window.

## 1. Introduction

Encouraged by recent advances in submillimeter very long baseline interferometry (VLBI)—including the Event Horizon Telescope's (EHT) groundbreaking results at 230 GHz in the first images of M87 and Sgr A* (EHT Collaboration 2019a, 2022a) and successful fringe detections at 345 GHz (Raymond et al. 2024)—we initiated an experiment to extend VLBI to even higher frequencies. The experiment conducted an observation at 685 GHz, the highest frequency ever attempted for ground-based VLBI, using the Atacama Large Millimeter/submillimeter Array (ALMA), the Atacama Pathfinder EXperiment (APEX), and the James Clerk Maxwell Telescope (JCMT). In addition, a benchmark VLBI test at 230 GHz was carried out to provide a lower-frequency reference, with participation from the Greenland Telescope (GLT; Chen et al. 2023).

The objective of this work was not only to achieve a technical milestone but also to evaluate the practical requirements—including instrumental readiness, calibration strategy, atmospheric conditions, and operational resources—necessary for sustained near-terahertz VLBI. By characterizing these constraints, this experiment serves as a pathfinder within a broader, long-term program aimed at extending high-frequency VLBI capabilities across multiple locations, including high-altitude sites in Greenland, Maunakea in Hawaii, and the Chajnantor Plateau in Chile. The present paper focuses on the 685 GHz experiment and its technical implications for ground-based near-terahertz VLBI performance.

With these technical considerations in mind, the scientific potential at 345 GHz and beyond is profound. Higher angular resolution could reveal time-variable emission structures on event-horizon scales in M87 and Sgr A*, such as orbiting hot spots or the foot points of relativistic jets, and the photon ring (Roelofs et al. 2019, Johnson et al. 2020). The strong scattering that blurs the Galactic Center at 230 GHz decreases sharply with frequency, raising the possibility of an unblurred, intrinsic image of Sgr A* (Johnson et al. 2018). Polarimetric observations at submillimeter and near-terahertz frequencies could probe magnetic fields in the near-horizon plasma, enabling tests of jet-launching mechanisms such as the Blandford–Znajek process (e.g., EHT Collaboration 2021, 2024). More broadly, multi-frequency imaging at the highest submillimeter bands would allow stringent tests of General Relativity, constraints on accretion physics in the strong-gravity regime, and perhaps even open paths to more speculative goals such as cosmological standard rulers or signatures of exotic horizon-scale physics (e.g., EHT Collaboration 2019c, 2022b). Additionally, higher angular resolution could resolve black hole images that cannot be reached at the current resolution (Pesce et al. 2021; Ramakrishnan et al. 2023; Zhao et al. 2024).

The challenges of submillimeter VLBI increase steeply with observing frequency. While VLBI at 230 GHz is now routine (EHT Collaboration 2019a, 2022a, 2024, 2025), success at 345 GHz and beyond demands exceptional weather, with both low atmospheric opacity and stable tropospheric phase (Raymond et al. 2024). Such conditions are rare, which makes higher-frequency VLBI both technically and logistically demanding.

Only a handful of ground-based observing facilities are situated at high, dry sites capable of observation at frequencies beyond 345 GHz. One such location is the Chajnantor plateau in Chile, site of both ALMA and the 12 m APEX telescope (e.g., Otárola et al. 2019). Another is Maunakea in Hawaii, site of the Submillimeter Array (SMA) and the JCMT (Naylor et al. 2000; Pardo et al. 2001a). Both of these locations are capable of achieving the requisite sky transparency to enable observations well above 600 GHz. Other sites, including the Antarctic plateau (e.g., Ishii et al. 2010; Tremblin et al. 2011; Shi et al. 2017) and the summit of the Greenland ice sheet (Matsushita et al. 2017a), offer exceptionally transparent and stable atmospheric windows into the terahertz regime. However, these locations currently lack the infrastructure required to support routine VLBI observations at such frequencies (Inoue et al. 2014).

The VLBI experiment described in this report was made possible by an opportunity to outfit the JCMT with a 690 GHz receiver (Bintley et al. 2024). The system combined a cryostat originally built for the Greenland Telescope (GLT) (Chen et al. 2023) with a new ALMA Band 9–type receiver cartridge developed by the NOVA laboratory in Groningen, the Netherlands, for the Large Latin American Millimeter Array (LLAMA). This upgrade provided JCMT with a dual-polarization, sideband-separating receiver, improving upon the legacy ALMA Band 9 double-sideband (DSB) design (Baryshev et al. 2015; Cortés et al. 2024).

Our goal in this 690 GHz experiment was to demonstrate the feasibility of VLBI at such high frequencies, thereby establishing a critical stepping stone toward future observations of the supermassive black holes Sgr A$^*$ and M87$^*$. This effort also lays the groundwork for expanding the range of sources accessible to black hole imaging science. In the following sections, we describe the preparation of the instruments, the frequency standards, ALMA Band 9 phasing efforts, the observations themselves, post-processing analyses, and a discussion of results leading to our conclusions.

## 2. ALMA Phasing System (APS)

The ALMA Phasing System (APS) enables ALMA to act as a single, beamformed element in global VLBI networks. By coherently summing signals from up to 61 antennas, it transforms ALMA into a phased array. This substantially enhances array sensitivity and extends VLBI baseline coverage, particularly by providing long north–south baselines that enhance the array's two-dimensional angular resolution.

A VLBI experiment at terahertz frequencies could not have been attempted without a functioning phasing system at ALMA Band 9. The ALMA Phasing Project (APP) implemented the first phasing capability more than a decade ago (Matthews et al. 2018). Initially, it targeted Bands 3 and 6, enabling ALMA's participation in the EHT and the Global Millimeter VLBI Array (GMVA; Krichbaum et al. 2008; Ros et al. 2024). Over time, the system's scope was extended to Bands 1, 3, 6, and 7, with added functionality for spectral-line phasing (Crew et al. 2023; Matthews & Crew 2024). The extension to Band 9 was initiated to exploit the site's optimal weather conditions and assess the feasibility of phased-array operation at such high frequencies. Software tests in early April 2024 confirmed that Band 9 phasing was technically viable. However, conditions in Band 9, as discussed in the later sections, are considerably less favorable than at Band 7, and a more sensitive version of the APS was required.

APS operates with two phase correction loops: a slow loop, updating every ∼16 s using internal cross-correlations, and a fast loop, applying ∼1 Hz corrections based on water vapor radiometer (WVR) data. Phase adjustments are implemented in real time via the ALMA Baseline (BL) correlator.

Usage of the APS in ALMA Band 9, however, posed unique challenges. Unlike the lower-frequency receivers, Band 9 employs a DSB architecture rather than a single-sideband (SSB) architecture. The intrinsic sensitivity penalty of a DSB receiver arises from noise from the unwanted sideband, which increases the effective system temperature. It raises questions about how such signals would be processed within ALMA's VLBI backend, which had previously been confirmed to work only for SSB operation (Crew & Matthews 2015). Analysis of the baseline correlator mechanisms suggested that VLBI should work as expected, but the system is complex enough that a true VLBI test was needed.

The sensitivity limitation is further compounded by the fact that the flux densities of extragalactic sources (including most commonly observed VLBI calibrators) tend to diminish at higher frequencies. To mitigate these challenges, a new phasing mechanism called the *delay-fix* was introduced—an upgrade that directly measures and corrects delays across the full 2 GHz baseband, replacing the earlier chunked (∼234 MHz) correction scheme. This approach reduces residual delay errors across the band and improves coherent summation efficiency.

The *delay-fix* applied full-band delay correction during the experiment. However, as will be shown in the next sections, the APS performance remained limited by signal-to-noise constraints at these frequencies, and the present dataset does not allow a clean separation of *delay-fix* performance from other contributing factors. Nevertheless, the implementation represents an important technical step toward robust high-frequency phased-array VLBI.

The legacy APS architecture is documented in early publications (Matthews et al. 2018; Goddi et al. 2019), while the new *delay-fix* is formally detailed in the APP2 Final Report (Matthews & Crew 2024).

## 3. APS Band 9 Testing

This section summarizes only those aspects relevant to the present VLBI experiment. The APS performance in Band 9 was evaluated through a series of test executions designed to assess the behavior of the newly implemented *delay-fix* mechanism under representative observing conditions. These tests were conducted as part of a targeted commissioning effort to determine whether phased-array operation at 690 GHz was technically viable. Three configurations were tested and compared: *delay-fix* enabled, legacy phasing, and *delay-fix* combined with an additional pointing calibration. Each execution included delay-calibration scans to compare measured instrumental delays with predictions from the BL correlator model. The level of agreement varied across runs, reflecting the sensitivity of high-frequency phasing to integration time, atmospheric stability, and calibration quality.

Phasing performance was quantified using internal coherence metrics. Configurations employing the *delay-fix* generally achieved higher coherence values than the legacy algorithm under comparable conditions. These tests indicate that Band 9 phasing is technically achievable when atmospheric stability and source brightness are sufficient.

At the same time, the tests highlight the intrinsic challenges of operation at these frequencies. Performance remains limited by reduced source flux densities, rapid atmospheric variability, and the tighter calibration tolerances imposed by wideband correction. In this sense, the system represents the current practical frontier of APS capability rather than a fully matured operational mode.

Further characterization will be required before routine high-frequency phased-array operation can be established. Ongoing developments at ALMA, including observatory-wide sensitivity upgrades (Carpenter et al. 2022), may contribute incrementally to improving operational robustness at the highest frequencies.

## 4. VLBI Planning, Preparation, and Observations

With the APS configuration established for Band 9 operation, phasing performance was initially evaluated during the 2024 EHT observing campaign with support from the ALMA VLBI team. The results indicated that phased-array operation at 690 GHz was technically plausible under favorable conditions, motivating the development of a dedicated VLBI experiment in collaboration with APEX and the JCMT. The experiment design addressed phasing performance, backend compatibility, and inter-station synchronization, and was finalized in October 2024.

A significant technical complication arose at JCMT when its VLBI frequency reference–normally derived from a hydrogen maser at the nearby Submillimeter Array–became unavailable due to required maintenance. To maintain schedule feasibility, an iodine-based optical atomic clock (Roslund et al. 2024) developed by Vector Atomic[31] (VA) was made available on loan for the trial. The clock was implemented as the frequency reference at JCMT and converted to a 10 MHz signal via a phase-locked local oscillator. Its stability was independently verified in this experiment through Allan deviation and PPS drift measurements, demonstrating performance comparable to a hydrogen maser on VLBI-relevant timescales. This validation enabled continuation of the experiment without reliance on the SMA maser.

Weather, always a critical factor in submillimeter VLBI, posed the final hurdle. Even at 230 GHz (Band 6) and 345 GHz (Band 7), excellent transparency and stable tropospheric conditions are required (EHT Collaboration 2019a, 2019b). By late October each year, optimal conditions at ALMA are increasingly rare, while those on Maunakea generally improve. In 2024 November, the window of opportunity opened: on the night of November 21, conditions were forecast to be acceptable, triggering the first 690 GHz VLBI test with ALMA-APEX-JCMT baseline network.

### 4.1. The Observing Targets

We selected three targets for this experiment. The primary target, J0423−0120, is an established ALMA Band 9 calibrator and has been regularly monitored. It was also included in a recent 345 GHz VLBI experiment (Raymond et al. 2024), where it remained unresolved on the ALMA–SMA baseline, indicating a compact structure on VLBI scales. During our observations, its flux density was measured to be ∼2.3 Jy at 690 GHz with ALMA. Near the time of the experiment, ALMA calibrator measurements report flux densities of ∼3.9 Jy at 233 GHz and ∼3.3 Jy at 343 GHz, indicating a negative spectral index with a mild decline toward higher frequencies (spectral index $\alpha \approx -0.3$, assuming $S_\nu \propto \nu^\alpha$). This spectral behavior suggests that the source remains bright into the submillimeter regime, supporting its suitability for VLBI fringe detection at 690 GHz.

The secondary target, J0433+0521, has been observed by ALMA primarily at Band 7 (∼343 GHz), with flux densities around ∼2 Jy near the time of our observations. No concurrent measurement at 690 GHz is available, and its spectral behavior toward higher frequencies is uncertain. As a result, its correlated flux density at Band 9 is difficult to predict, making it a less favorable target for fringe detection at 690 GHz compared to J0423−0120.

At the time of the experiment, these two sources were the only viable candidates identified for fringe detection given the planned observation time. Both are relatively bright in submillimeter wavelengths and are located such that they could be observed from both ALMA/APEX and JCMT at elevations exceeding 30°, ensuring favorable atmospheric transmission.

In addition, the Becklin–Neugebauer (BN) object in Orion (ORIBN) was selected for dedicated scans to verify the Band 9 spectral-line tuning. ORIBN is a compact and bright submillimeter source in the Orion KL region with strong molecular line emission, making it suitable for tuning validation at 690 GHz.

For the primary and secondary targets, we also scheduled a Band 6 (230 GHz) session as a VLBI benchmark test of overall system, since observations in this band are well established. The GLT was included for this session to expand the baseline coverage.

### 4.2. Frequency Standard

In this experiment, each participating station operated with its own frequency standard. ALMA, APEX, and GLT were referenced to hydrogen masers, the conventional standard for VLBI applications. In contrast, JCMT was referenced to an optically stabilized 10 MHz signal derived from a VA iodine clock, as introduced earlier.

Hydrogen masers provide excellent short-term phase stability and are widely adopted in VLBI systems. In contrast, an iodine clock is a laser-based frequency standard in which a stabilized laser is locked to a narrow molecular iodine transition and its optical stability is transferred to microwave frequencies via optical frequency division. Owing to the high optical carrier frequency, such systems achieve low phase noise and strong short-term fractional frequency stability. Recent transportable iodine clocks have demonstrated short-term instabilities of order $5 \times 10^{-14}/\sqrt{\tau}$, where $\tau$ is the averaging time in seconds, reaching fractional stability below $10^{-14}$ over multi-day averaging intervals, with performance comparable to active hydrogen masers in compact and field-deployable architectures (Roslund et al. 2024).

[31] https://vectoratomic.com/ (accessed 2025).

Since JCMT employed a precision optical clock, marking a pioneering application in VLBI, the clock's stability required careful verification. Prior to the observations, two VA iodine clocks were installed in the SMA hydrogen maser room, from which JCMT derived its frequency standard. The primary VA unit provided two stable outputs. A 10 MHz reference was sent directly to a low-phase-noise Keysight E8257D synthesizer, which served as the primary LO reference for JCMT. A second output, a 100 MHz reference, was combined with the primary LO and transmitted over optical fiber to JCMT. Because the optical system is specified for transmission at 100 MHz or higher and cannot directly carry a 10 MHz signal, a local 10 MHz oscillator at JCMT was phase-locked to the received 100 MHz signal and used as the master reference for the JCMT backend systems. (Kubo et al. 2018; Chen et al. 2023).

The second VA unit was configured for independent evaluation against both the primary and a Rakon[32] oscillator in the same location. Stability was additionally assessed using JCMT's Global Positioning System (GPS) receiver, confirming short- and long-term performance with residual drift constrained to within 7 fs $s^{-1}$, well within VLBI requirements. On short timescales, the VA clock exhibited stability comparable to, and in some regimes exceeding, that of the SMA hydrogen maser it replaced.

Allan deviation and drift rate were also measured for the frequency standards at all participating stations. The frequency offsets were 6.1 fs $s^{-1}$ for ALMA, 1041.4 fs $s^{-1}$ for APEX, and 6.8 fs $s^{-1}$ for JCMT. The comparatively large drift observed at APEX was stable over the observing interval and is removed during VLBI correlation.

We measured the differential Allan deviation of the 10 MHz reference signals shortly before the VLBI experiment. The comparison between the Rakon oscillator and the VA primary output reflects the stability of the optically derived 10 MHz reference relative to a high-quality crystal oscillator. The results show that the stability at averaging times $\tau \gtrsim 2$ s is dominated by the Rakon oven-controlled crystal oscillator (OCXO).

A separate comparison between two optically derived 10 MHz outputs indicates a fractional stability of approximately 5.0e-14 at $\tau = 1$ s, 1.5e-14 at $\tau = 10$ s, and 1.5e-15 at $\tau = 100$ s.

Overall, the frequency standards deployed in this experiment satisfied the phase stability requirements for 690 GHz VLBI. The hydrogen masers at ALMA, APEX have conventional and well-characterized performance, while the optically stabilized reference at JCMT demonstrated short-term stability comparable to maser systems after independent verification.

[32] https://www.rakon.com/

### *4.3. RF Frequency and LO*

The selection of the observing frequency was based on considerations of the optimal system temperature for ALMA Band 9 and the desire to cover the CO(6–5) spectral line for system verification, including frequency tuning and signal detection at VLBI stations. Thus, we set the primary local oscillator (LO) frequency to 685 GHz to capture four sub-bands, centered at 678 GHz, 680 GHz, 690 GHz, and 692 GHz, respectively, each with a 2 GHz bandwidth. The JCMT and APEX use dual-polarization, 2SB receivers (Belitsky et al 2018; Bintley et al 2024), and the data were processed with the EHT hardware and recorders as described in (EHT Collaboration 2019c). Since ALMA Band 9 is DSB, it requires further steps to extract the desired signals.

For standard interferometry observations in Bands 9 and 10, ALMA uses 90-degree Walsh (Thompson et al. 2017, hereafter TMS) phase-switching sequences to separate DSB signals into single-sideband (SSB) data within the ALMA baseline (BL) correlator. This setup allows the four basebands (the four IFs from a dual-pol, 2SB receiver) to produce eight distinct SSB spectral windows in ALMA Bands 9 and 10 (Cortés et al. 2024). However, this setup involves processes outside the APS domain and is unusable for phasing and VLBI. The purpose of the APS is to provide a combined, coherent signal for recording, and the 90° Walsh-switching process completely scrambles the signal that is ultimately recorded.

Thus, during the APS operation, the 90° Walsh-switching must be turned off, and the DSB-to-SSB signal conversion is also disabled. Since ALMA Band 9 remains intrinsically DSB at the frontend, it provides a special function to extract the required sideband while suppressing the other sideband using LO offsetting (the so-called "signal" sideband in Cortés et al. 2024, Section 6.3).

ALMA enacts a second Walsh sequence series that applies a 180° phase shift to the signals from each antenna, and is demodulated by a sign change within the BL correlator; as a result, the wanted signals correlate, and the unwanted signals are canceled out. This rejects spurious signals generated in the system between the receiver and the BL correlator. Such a Walsh-switching scheme is used in the ALMA VLBI observations, and has been shown to be compatible with the summing operation: the Walsh-switching phase shift insertion and removal all happen before the signal is summed and thus there are no issues (as demonstrated by the EHT results).[33]

[33] We did not investigate in detail, but the combination of APS processing and 180° Walsh-switching apparently renders one of the sidebands unusable for the VLBI correlation (as would normally be expected in VLBI).

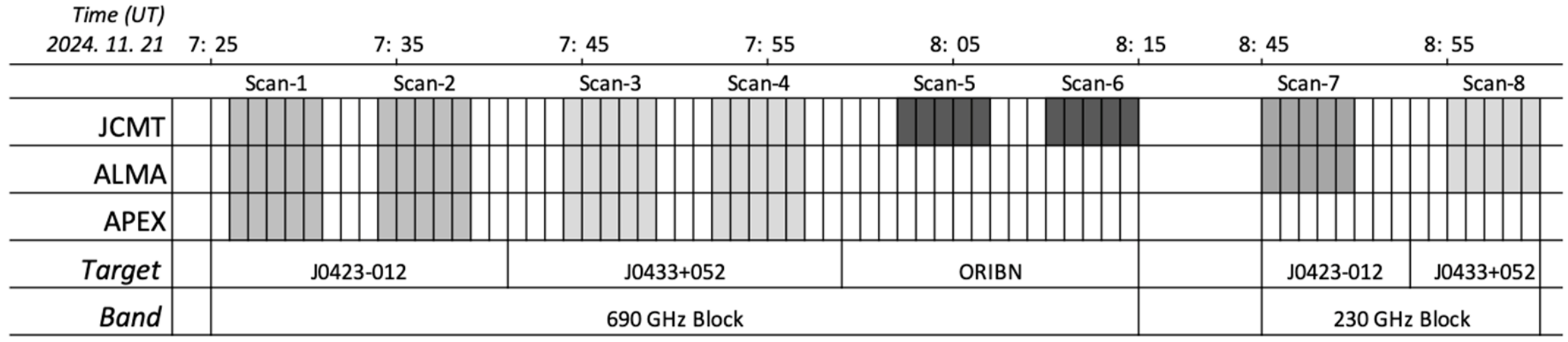

**Figure 1.** Observation schedule of the 690 GHz VLBI experiment conducted on 2024 November 21.

**Table 1**
A List of the ALMA 12 m Antennas used in the Observing Array

| | | | | | | | | | | | |
|---|---|---|---|---|---|---|---|---|---|---|---|
| DA45 | 0.000 | DA63 | 16.744 | DA57 | 18.338 | DA41 | 18.585 | DA51 | 18.654 | DA48 | 26.892 |
| DV20 | 34.560 | DV17 | 35.128 | DV16 | 36.798 | DA64 | 40.183 | DV05 | 46.417 | DV22 | 52.220 |
| DV10 | 54.774 | DA43 | 55.097 | DV08 | 55.950 | DV06 | 56.162 | DV04 | 57.233 | DA52 | 57.457 |
| DA60 | 59.382 | DV15 | 60.076 | DV01 | 61.195 | DV02 | 62.243 | DA47 | 62.617 | DV13 | 66.585 |
| DV19 | 67.228 | DA59 | 69.223 | DV24 | 72.651 | DV09 | 73.041 | DA54 | 75.390 | DA50 | 77.273 |
| DA58 | 79.816 | DV11 | 80.125 | DV03 | 83.955 | DA56 | 84.007 | DA42 | 84.999 | DA65 | 89.881 |
| DA46 | 94.489 | DV25 | 99.006 | DA49 | 100.100 | DA53 | 134.512 | DV21 | 136.235 | DV12 | 139.598 |
| DA61 | 152.884 | DV23 | 153.718 | DA62 | 163.005 | DV14 | 183.934 | | | | |

**Note**. The "sum" antenna, not shown in the table, exists only virtually, and its signal is presented to the BL correlator as if it came from the 7 m antenna CM12 but located at the array reference position. Numbers indicate distances in meters to the reference antenna DA45.

## 5. 2024 November 21 Observation

The experiment was allocated five hours, including time for the prior phasing tests and a 90 minute VLBI observing block. As in standard EHT operations, a week-long trigger window was assigned to identify suitable weather simultaneously at Maunakea and the Chajnantor plateau. Within this window, only 2024 November 21 provided acceptable conditions at both sites, and the experiment was triggered on 20 November.

On 2024 November 21 (UT), weather conditions at ALMA/APEX and JCMT were excellent, with precipitable water vapor (PWV) values of 0.46 mm and 1.3 mm, respectively, and negligible wind at both sites. JCMT participated fully in the VLBI experiment, recording all scheduled scans at the designated tunings. APEX operated nominally, aside from a brief initial delay.

The GLT was scheduled to participate in this observation, with receiver and backend systems configured analogously to the 230 GHz EHT setup (Chen et al. 2023). However, adverse weather conditions at the GLT site prevented usable data acquisition; no instrumental issues were identified.

Figure 1 summarizes the observing schedule for ALMA, JCMT, and APEX. All three stations participated in the 690 GHz observations during the first four scans. Scans 5 and 6 were used to verify the spectral-line tuning. Scans 7 and 8 were conducted at 230 GHz with ALMA and JCMT; APEX did not participate in the 230 GHz block.

Both the primary and secondary targets were observed throughout the schedule, even though the secondary target did not provide sufficient correlated flux density at Band 9 for robust phasing with the available 2 GHz bandwidth.

For the Band 9 observations, the ALMA Phasing System (APS) coherently combined all available 12 m antennas within 200 m of DA45. The most distant participating antenna from DA45 was DV14 at 183.9 m. Antenna DV01 (61.2 m from DA45) was excluded from the phased sum and used as an independent comparison antenna to evaluate phasing performance. The full antenna list is given in Table 1.

Within the APS workflow, the phased output of the array is represented as a virtual "sum" antenna. This sum signal has no physical location and is formed by coherently combining the voltages from the participating antennas. For correlation and analysis, the sum signal is injected into the correlator as a single antenna input and treated as an ordinary antenna. In the data products, it inherits the name of the antenna to which it is logically attached—CM12, a 7 m antenna not otherwise part of the 12 m array for these observations. By convention, the sum antenna is assigned to the array reference position, and distances indicated in Table 1 are measured relative to DA45.

## 6. Results and Analysis

Despite the favorable weather conditions and nominal system configuration, a significant issue became apparent at the start of the 690 GHz observations: the ALMA Phasing

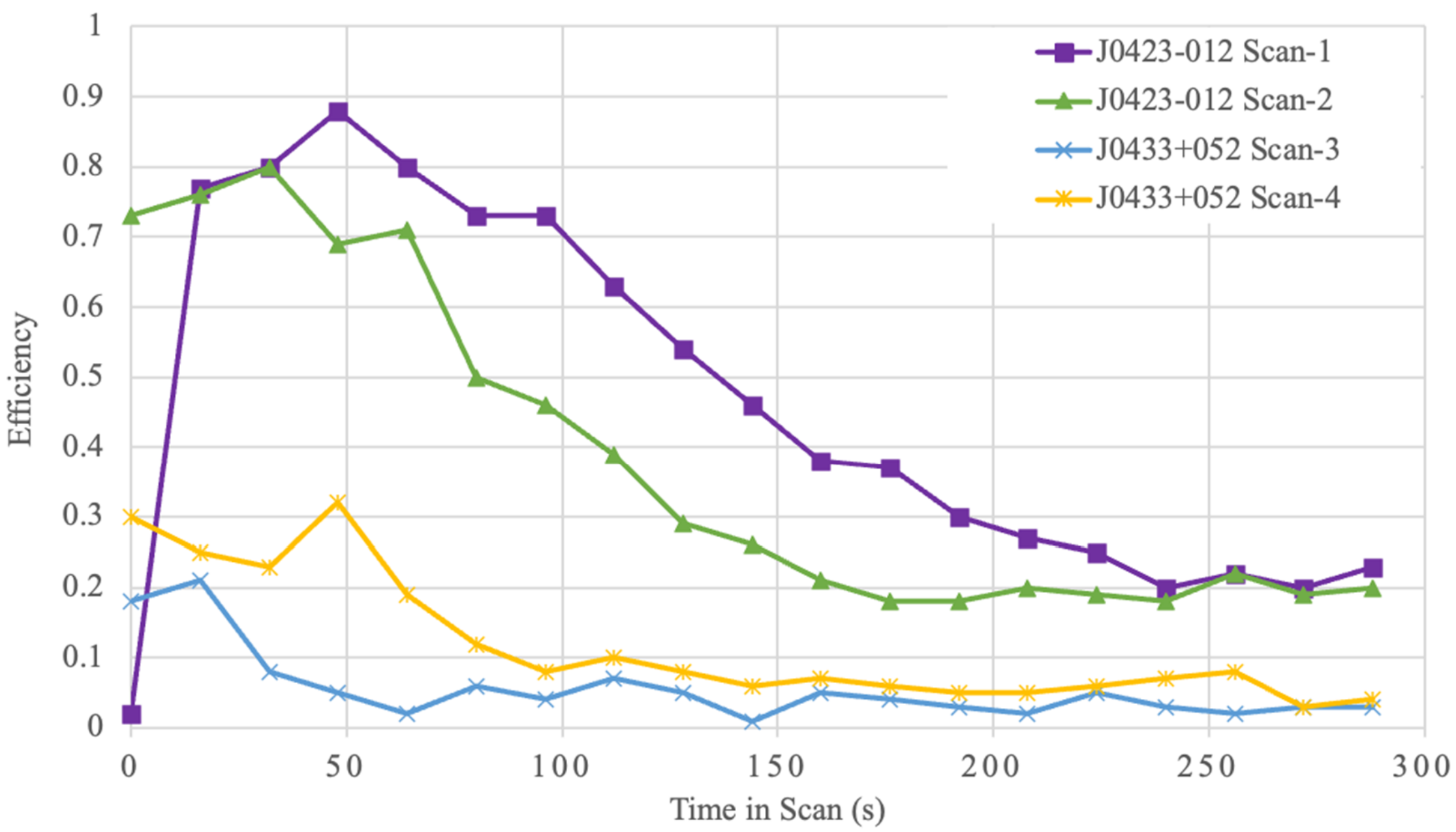

**Figure 2.** Measured APS Efficiency—This shows the as-reported values for the phasing efficiency (vertical) axis during the ALMA Band 9 observation as a function of time during the scan. The first two scans (purple and green) were on the brighter source and began with 70% or better efficiency and deteriorated after that. The second pair (blue and orange) was on the weaker source and was never well phased. Note that all four scans show a similar pattern; they start at a relatively higher phasing efficiency and slowly degrade toward the end of the scans. The APS was operated to retain phases through the mechanisms of the ∼1 Hz WVR corrections, so that only the very first scan on the first source (purple line at 0) was completely unphased. For the second pair (blue and orange), the reported efficiency is essentially a random result after 100 s. In comparison, the 230 GHz scans all maintained a phasing efficiency (not shown here) above 90% throughout the 300 s scan.

System (APS) did not maintain the expected phasing stability. Figure 2 shows the near-real-time phasing efficiency of the four scans in the 690 GHz block. For the first two scans on the primary targets, the phasing efficiency initially started reasonably well, as expected due to the excellent weather conditions, but subsequently degraded towards a level of 30%.

The two scans on the secondary target exhibited a similar behavior. Although the initial efficiencies were modest, they declined rapidly during each scan. The similar pattern in all the scans suggests that the APS could only sustain coherent phasing for about one minute before losing effective phase summation.

Such degradation has not been observed at lower frequencies, and additional testing would be required to fully understand its origin. Given that new atmospheric delay solutions are derived every 16 s, the cadence of the delay model alone is unlikely to account for the observed behavior at Band 9. Instead, the degradation may reflect differences in APS fringe-fitting and phasing behavior under these extreme conditions, and the current system may be adequate only for shorter scans at this frequency (e.g., at most 90 s as we might judge from Figure 2).

In contrast, the 230 GHz scans all maintained a phasing efficiency above 90% throughout the 300 s scan. Such performance is routinely achieved at the lower ALMA bands under appropriate weather conditions and elevations above ∼40° (e.g., Matthews et al. 2018), confirming that the APS was functioning nominally at 230 GHz.

The implications of the phasing degradation in Scans 1 and 2 for correlation and fringe detection will be discussed in a later section.

### 6.1. Correlation

Correlation was performed at MIT Haystack using *DiFX* (Deller et al. 2011), version 2.7.1, with subsequent analysis carried out using the Haystack Observatory Postprocessing System (HOPS[34]) package. The results are summarized below.

The Band-9 data required several rounds of correlation because early attempts were affected by inconsistent metadata, timing mismatches, and intervals in which the array was not coherently phased. The key breakthrough came only after the Band-6 (∼230 GHz) reference scans were successfully correlated. Their strong, stable fringes demonstrated that the overall VLBI network was functioning correctly and that the observing setup was configured as intended. This established a reliable foundation for reassessing the higher-frequency data.

[34] https://www.haystack.mit.edu/haystack-observatory-postprocessing-system-hops

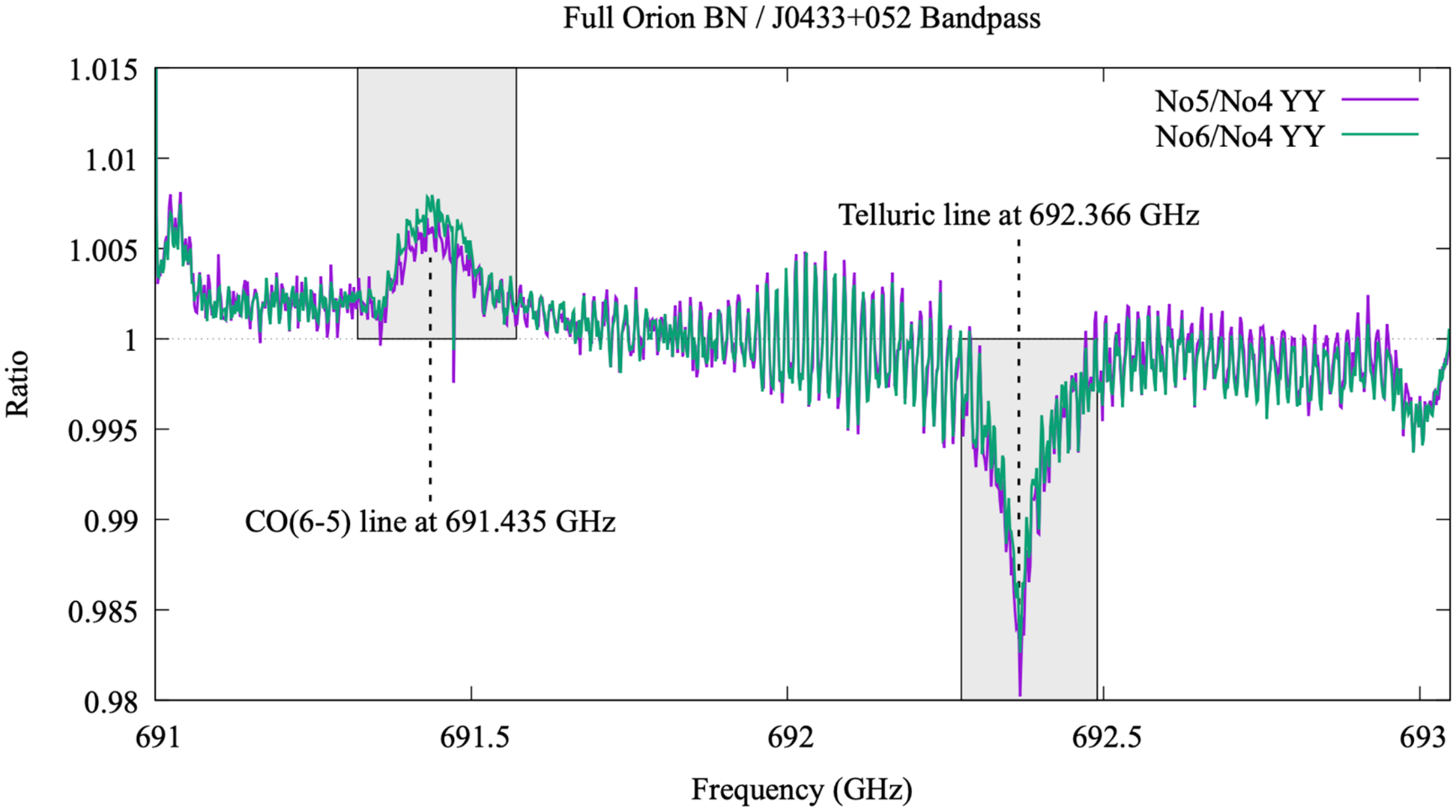

**Figure 3.** Band-9 spectral tuning check using Orion BN. Shown are the spectral ratios from Scan 5 and Scan 6, each referenced to Scan 4 (J0433+0521). Only the YY polarization is shown (see text for details). The CO (6–5) line is detected at the expected rest frequency of 691.435 GHz, and a prominent absorption feature near 692.366 GHz is present, consistent with a telluric water-vapor absorption band in this frequency range. Shaded regions are included solely to guide the eye to these spectral features. A representative 20 s segment was used for this diagnostic to limit correlator usage. The persistence of both the astrophysical CO emission and the atmospheric absorption feature in the calibrated spectra confirms the correctness of the JCMT Band-9 frequency tuning and IF configuration. Spectral difference plots show consistent behavior.

The Band-6 observation (Scan 7 and Scan 8) showed a wide range of high signal-to-noise ratio (SNR) values across all four polarization products. They revealed that ALMA and JCMT shared almost identical clock and delay-rate solutions, an essential indicator that the stations were properly aligned in both time and frequency. The subband spectra were clean and fully populated, and the delay and rate peaks were sharply defined in every polarization. These results confirmed that the Band-6 data behaved as expected for bright calibrators, providing the confidence needed to reprocess Band-9 using only the valid overlap windows, as shown in the next few sections.

Both Scan 5 and Scan 6 clearly detect the CO (6–5) line at 691.435 GHz, as expected for the Orion BN target. In addition, a prominent telluric absorption feature is observed near 692.366 GHz, consistent with the strong atmospheric water-vapor band in this frequency range (Naylor et al. 2000). The spectra were extracted from the JCMT VLBI autocorrelation data produced by the correlator. In a manner analogous to sky switching, Scan 4 served as a reference background against which Scans 5 and 6 were processed using both ratio and subtraction to highlight weak spectral structure.

Although several minutes of data were obtained on Orion BN, only a representative 20 s segment was processed for this spectral check in order to limit correlator usage. As a result, the intrinsic spectral signals are weak, and the Orion BN spectra are examined relative to Scan 4 (the J0433+0521 bandpass calibrator) using both division and subtraction. For clarity, only the YY polarization is shown in Figure 3; the corresponding XX results are similar. In both representations, the CO emission and the telluric absorption remain clearly visible. These two independent spectral anchors—one astrophysical and one atmospheric—demonstrate that the observed features are not instrumental artifacts and confirm that the JCMT Band-9 frequency tuning and IF configuration were correctly set for this experiment.

The Band-9 fringe test used the first three scans (Scan 1, 2 and 3) with the two LO tunings[35] (b2 $\approx$ 679 GHz and b3 $\approx$ 691 GHz). Short 70 s test correlations on Scan 2 produced clear detections across all polarization products, confirming that both tunings generated real Band-9 fringes. Scan 1 did *not* yield a reliable detection in either tuning because APEX began recording roughly 40 s late, leaving too little usable overlap for a stable fringe. Scan 3 also produced no detection, consistent with its known data-quality issues and missing polarization channels. Thus, Scan 2 provides the sole robust Band-9 detection, demonstrating that the frequency setup and correlation pipeline were functioning correctly.

[35] Multiple and sometimes inconsistent nomenclatures exist across ALMA and the EHT regarding high-frequency tuning. At ALMA, the Band-9 system and baseline correlator produce two basebands, BB_2 and BB_3. In VLBI processing, each baseband is treated as a single spectral window (typically labeled SW-01), and for convenience the VLBI recordings corresponding to BB_2 and BB_3 are commonly referred to as "b2" and "b3." In standard EHT observations, all four ALMA basebands are used, but in this experiment only BB_2 and BB_3 were correlated. At JCMT and APEX, the local oscillator (LO) was tuned to 685 GHz, producing corresponding IF placements that map naturally onto the ALMA BB_2/b2 and BB_3/b3 nomenclature used here.

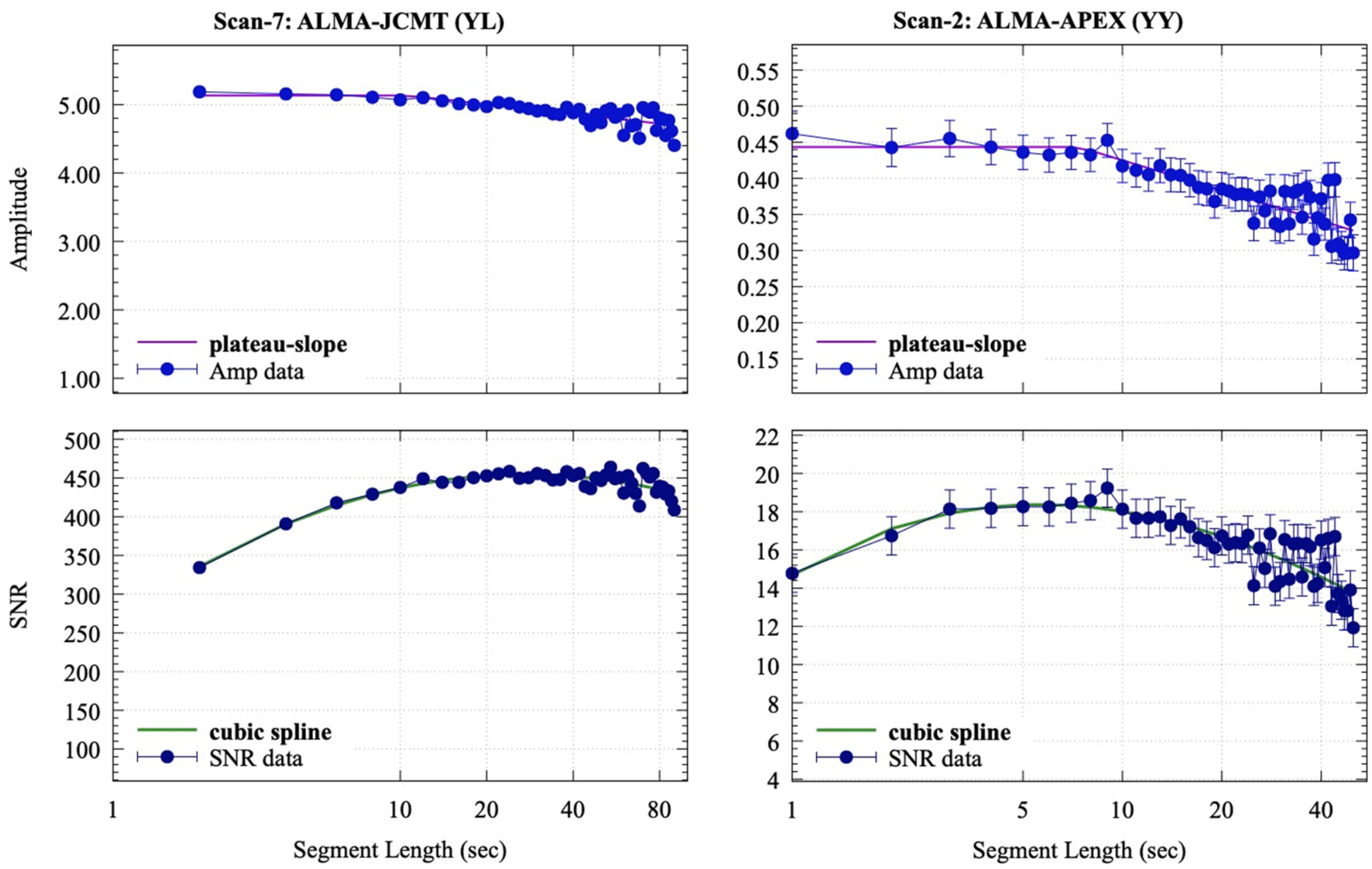

**Figure 4.** Fringe amplitude and SNR as a function of solution interval for VLBI detections at 230 GHz (left) and 690 GHz (right). In each case, the upper panel shows the correlated fringe amplitude as a function of segment length, while the lower panel shows the corresponding fringe SNR. The plateau–slope guide lines shown in the amplitude panels are intended only to aid visual interpretation of the coherence behavior. The curves overlaid on the SNR panels are cubic-spline interpolations used to guide the eye. The 230 GHz data exhibit the expected behavior for partially coherent averaging, with approximately constant short-timescale amplitude followed by gradual coherence loss at longer integrations. In this case, the statistical uncertainties are smaller than the plotted symbols. In contrast, the 690 GHz data show a more rapid decrease in correlated amplitude and greater scatter, indicating stronger short-timescale phase fluctuations. Nevertheless, useful fringe SNR is maintained over substantially longer integration intervals than implied by the formal amplitude coherence estimate alone.

The fringe-detection analysis was performed using the *cohfit* utility within the HOPS analysis package. While *cohfit* itself is an internal HOPS tool,[36] it implements statistical fringe-detection and coherence concepts developed for VLBI data analysis (e.g., Rogers et al. 1995) within the Mark 4/HOPS framework (Whitney et al. 2004). This analysis examines the fringe-fitting results as a function of solution interval, defined here as the time span over which a single fringe solution is estimated. The visibility data are segmented into intervals of different durations, and fringe solutions are obtained over a range of solution intervals. As expected, solutions obtained over shorter intervals remain stable, whereas longer ones show a gradual loss of signal, reflected in reduced fringe amplitude and SNR after incoherent averaging over the segmented intervals. Representative examples of the solution-interval dependence behavior at 230 GHz and 690 GHz are shown in Figure 4.

At 230 GHz, the ALMA–JCMT data show robust fringe recovery over the tested range of solution intervals. The fringe amplitude remains approximately stable over short to intermediate intervals, while the SNR increases only slowly with integration time. The SNR growth does not follow the $t^{\frac{1}{2}}$ ($t$ as the integration time) scaling expected for ideal coherent averaging, but is more consistent with the segmented, incoherent-averaging approach used in *cohfit,* in which fringe strength can still be accumulated when residual phase fluctuations limit fully coherent averaging (Rogers et al. 1995).

The 230 GHz ALMA–JCMT detections do not show a sharp amplitude cutoff over the tested segment lengths. The strong *fourfit* detections are also obtained over longer usable portions of the scans. Although the nominal scan duration is approximately 300 s, only about 100 s of data is effectively usable because of phasing stability and data-quality limitations, and the 230 GHz fringe solutions remain robust across this usable interval.

[36] HOPS is in transition from v3, generally released as part of the DiFX correlator software suite, to a revised v4 which will be released separately. The HOPS software used in this analysis will is released as version 3.26 and available at https://www.haystack.mit.edu/haystack-observatory-postprocessing-system-hops.

In contrast, the 690 GHz data show evidence for appreciable fringe degradation on timescales of order 10 s or longer. Although the nominal scan duration is similar, the usable data are more limited, and the fringe-search intervals are typically restricted to approximately 70 s because of the performance of the ALMA Phasing System. Within these intervals, the data exhibit significant scatter, and the inferred fringe-detection timescale depends on the choice of solution interval and analysis method. The SNR diagnostic reaches its maximum at solution intervals of approximately 5–10 s and decreases by about 10% at 20 s, with large losses toward longer integrations. Nevertheless, in practice, fringe solutions remain recoverable and yield consistent detections for integration times of order 20 s. We therefore use 20 s as a nominal solution interval for sensitivity estimation of the ALMA–APEX 690 GHz fringe detection.

We note that the ALMA–APEX baseline is much shorter than a typical intercontinental VLBI baseline. Because the two sites are separated by only a few kilometers, atmospheric phase fluctuations sampled by the two stations may remain partially correlated (Matsushita et al. 2017b). As a result, we treat the ALMA–APEX behavior as a special case and primarily as an empirical diagnostic of the detected 690 GHz fringe and of the experiment-specific fringe stability, rather than as a general measurement of the atmospheric coherence time for long-baseline 690 GHz VLBI.

It is therefore useful to estimate a separate reference solution interval for the ALMA–JCMT baseline at 690 GHz. Specifically, we estimate the solution interval that would be expected to produce the highest SNR in a comparable fringe-detection analysis. Rather than directly adopting the ALMA–APEX value, we use the 230 GHz ALMA–JCMT data as the empirical reference because they sample the same long baseline under similar observing conditions. In Figure 4, the SNR curve for the 230 GHz detection peaks at a solution interval of approximately 30 s. If atmospheric fluctuations are the primary cause of the SNR degradation, the associated non-dispersive path-length fluctuations imply residual phase fluctuations that scale approximately linearly with observing frequency (e.g., TMS; Carilli & Holdaway 1999). As a simple approximation, we therefore scale the reference solution interval inversely with frequency. The corresponding interval at 690 GHz is reduced by about a factor of three, suggesting an approximate solution interval of order 10 s for the ALMA–JCMT baseline in a similar fringe-detection analysis.

In the final check, a full-length Band-9 correlation was carried out on the ALMA–JCMT baseline using Scan 2 to determine whether a detectable high-frequency fringe was present. This attempt yielded no fringe, indicating that in this particular scan, the ALMA–JCMT pair did not produce a measurable Band-9 signal. This result does not reflect a failure of the correlator; rather, it simply shows that, for this dataset, only the ALMA–APEX baselines produced usable Band-9 detections.

To place these results in a more general and consistent context, Table 2 presents a subset of *fourfit* output parameters most relevant to fringe detection. Additional diagnostic quantities produced by *fourfit* are not listed. The table lists the *fourfit* fringe quality (Qcode), signal-to-noise ratio (SNR), probability of false detection (PFD), fitted multi-band delay, fringe rate, and effective integration time for each scan and polarization product, together with a qualitative classification of the fringe outcome (detected, marginal, or non-detection). This classification is based not solely on statistical significance, but also on the physical consistency of the fringe solution, including stability of the delay–rate peak across parallel-hand polarizations and comparison between adjacent scans.

Figure 5 illustrate representative cases using *fourfit* (Rogers et al. 1995) fringe diagnostics. The figure contrasts a robust, high-coherence detection on the ALMA–JCMT baseline at 214 GHz with a weaker but statistically significant detection on the ALMA–APEX baseline at 690 GHz, highlighting the pronounced impact of atmospheric phase stability and coherence time on high-frequency VLBI fringe recovery.

In summary, the Band-6 reference scans confirmed correct timing, frequency alignment, and correlator configuration. For Band 9, a robust fringe detection was obtained on the ALMA–APEX baseline in Scan 2 under the 691 GHz tuning, while no reliable detection was recovered in Scan 1 due to limited temporal overlap or in Scan 3 due to known data-quality issues. No Band-9 fringe was detected on the ALMA–JCMT baseline. These results establish the correlation outcome that forms the basis for the subsequent analysis.

### 6.2. Sensitivity Estimation

Due to the limited allotted ALMA time allocated to this experiment, no dedicated calibration scans were included. Thus, sensitivities were estimated from station technical parameters. Table 3 lists the system temperatures ($T_{sys}$) and the system equivalent flux density (SEFD) for each station during Scans 1 and 2. The system temperatures of ALMA and JCMT were measured during the observation. Their SEFDs are derived using radiometer theory, namely,

$$\mathrm{SEFD} = \frac{2kT_{sys}}{A_e}, \tag{1}$$

where $k$ is the Boltzmann constant and $A_e$ the effective collecting area. In this work, the aperture efficiencies for Band 9 are taken as 0.43 (Table 9.3, Cortés et al. 2024) and 0.32[37] for ALMA and JCMT, respectively. The APEX SEFD was extracted from a $T_{\mathrm{sys}}$ measurement using hot, cold, and

[37] http://docs.eao.hawaii.edu/JCMT/OVERVIEW/tel_overview/tel_overview.html#foot408

**Table 2**
Examples of Fringe-detection Statistics for the ALMA–JCMT (AM) and ALMA–APEX (AX) Baselines

| Scan _ID | Baseline | Freq_GHz | Pol | Qcode | SNR | PFD | SBD (ns) | MBD (ns) | Fringe Rate (Hz) | FSI (s) | Note |
|---|---|---|---|---|---|---|---|---|---|---|---|
| 7 | AM | 214.163 | XL | 5 | 321 | 0.0 | −39.8 | 1.54 | −0.174 | 300 | Detected |
| 7 | AM | 214.163 | YR | 5 | 238 | 0.0 | −41.6 | −0.08 | −0.174 | 300 | Detected |
| 7 | AM | 226.163 | XR | 5 | 185 | 0.0 | −25.6 | −1.23 | −0.185 | 300 | Detected |
| 7 | AM | 226.163 | YL | 5 | 358 | 0.0 | −25.2 | −0.76 | −0.185 | 300 | Detected |
| 8 | AM | 214.163 | XL | 6 | 276 | 0.0 | −40.3 | 1.10 | −0.173 | 300 | Detected |
| 8 | AM | 214.163 | YR | 6 | 270 | 0.0 | −42.0 | −0.52 | −0.173 | 300 | Detected |
| 8 | AM | 226.163 | XR | 6 | 302 | 0.0 | −26.0 | −1.67 | −0.182 | 300 | Detected |
| 8 | AM | 226.163 | YL | 6 | 312 | 0.0 | −25.6 | −1.19 | −0.182 | 300 | Detected |
| 1 | AX | 679.063 | XX | 0 | 4.8 | 0.7 | −175.0 | −1.29 | −0.345 | 70 | Non-detection |
| 1 | AX | 679.063 | YY | 0 | 4.9 | 0.6 | −31.7 | 1.15 | −0.383 | 70 | Non-detection |
| 1 | AX | 689.121 | XX | 9 | 7.3 | 0.0 | 27.3 | −0.84 | −0.294 | 70 | Marginal |
| 1 | AX | 689.121 | YY | 9 | 7.5 | 0.0 | 6.5 | −0.02 | −0.283 | 70 | Marginal |
| 2 | AX | 679.063 | XX | 9 | 10.4 | 0.0 | 39.0 | −5.38 | −0.292 | 70 | Marginal |
| 2 | AX | 679.063 | YY | 9 | 10.8 | 0.0 | 33.4 | 6.24 | −0.290 | 70 | Marginal |
| 2 | AX | 689.121 | XX | 9 | 12.9 | 0.0 | 25.7 | −1.14 | −0.293 | 70 | Detected |
| 2 | AX | 689.121 | YY | 9 | 12.6 | 0.0 | 9.3 | −0.04 | −0.292 | 70 | Detected |
| 1 | AM | 679.063 | XX | 0 | 4.7 | 0.8 | −166 | −1.74 | −0.238 | 70 | Non-detection |
| 1 | AM | 679.063 | YY | 0 | 4.8 | 0.8 | 101 | 0.66 | −0.386 | 70 | Non-detection |
| 2 | AM | 689.121 | XX | 0 | 5.9 | 0.9 | −1658 | 6.62 | −0.001 | 196 | Non-detection |
| 2 | AM | 689.121 | YY | 0 | 6.1 | 0.6 | 2268 | −4.45 | 0.501 | 200 | Non-detection |

**Note.** For each scan and polarization product, we list the *fourfit* fringe quality (Qcode), signal-to-noise ratio (SNR), probability of false detection (PFD), fitted multi-band delay (MBD), fringe rate, and fringe search interval (FSI). The listed interval corresponds to the portion of each scan with usable data included in the fringe analysis. Further details on the definition of the FSI and its relation to scan duration, usable data, and coherence time are provided in the text. Fringe classifications ("Detected," "Marginal," or "Non-detection") are based on SNR and PFD, together with the physical consistency of the fringe solution, including stability of delay and rate across parallel-hand polarizations and, where applicable, between adjacent scans. Single-band delay diagnostics were examined but are not used as primary criteria. At Band 9, isolated low-PFD candidates lacking robust coherence or repeatability are conservatively classified as marginal. All scans target J0423−0120 except Scan-8, which observed J0433+0521. Shaded rows indicate detected or marginal scans.

sky loads on 2025 April 5, adopting the APEX-specific conversion factor of 62 Jy $K^{-1}$ as documented on the APEX website.[38]

To derive the phased-ALMA SEFD, several real-world effects reduce the ideal sensitivity. The ALMA front end digitizes the signal with 3-bit quantization, resulting in a quantization efficiency of 0.96. The dominant degradation arises from the phasing process itself. Matthews et al. (2018, Section 8.2) reported an end-to-end phasing efficiency of approximately 61% under typical conditions. This loss reflects a combination of effects, including residual atmospheric phase noise, non-optimal antenna weighting, uncorrected baseband delay slopes, and truncation associated with the formation of the summed VLBI signal. With subsequent improvements to delay correction, phasing efficiencies of ∼70% have been demonstrated.

[38] https://www.apex-telescope.org/telescope/efficiency/?yearBy=2024

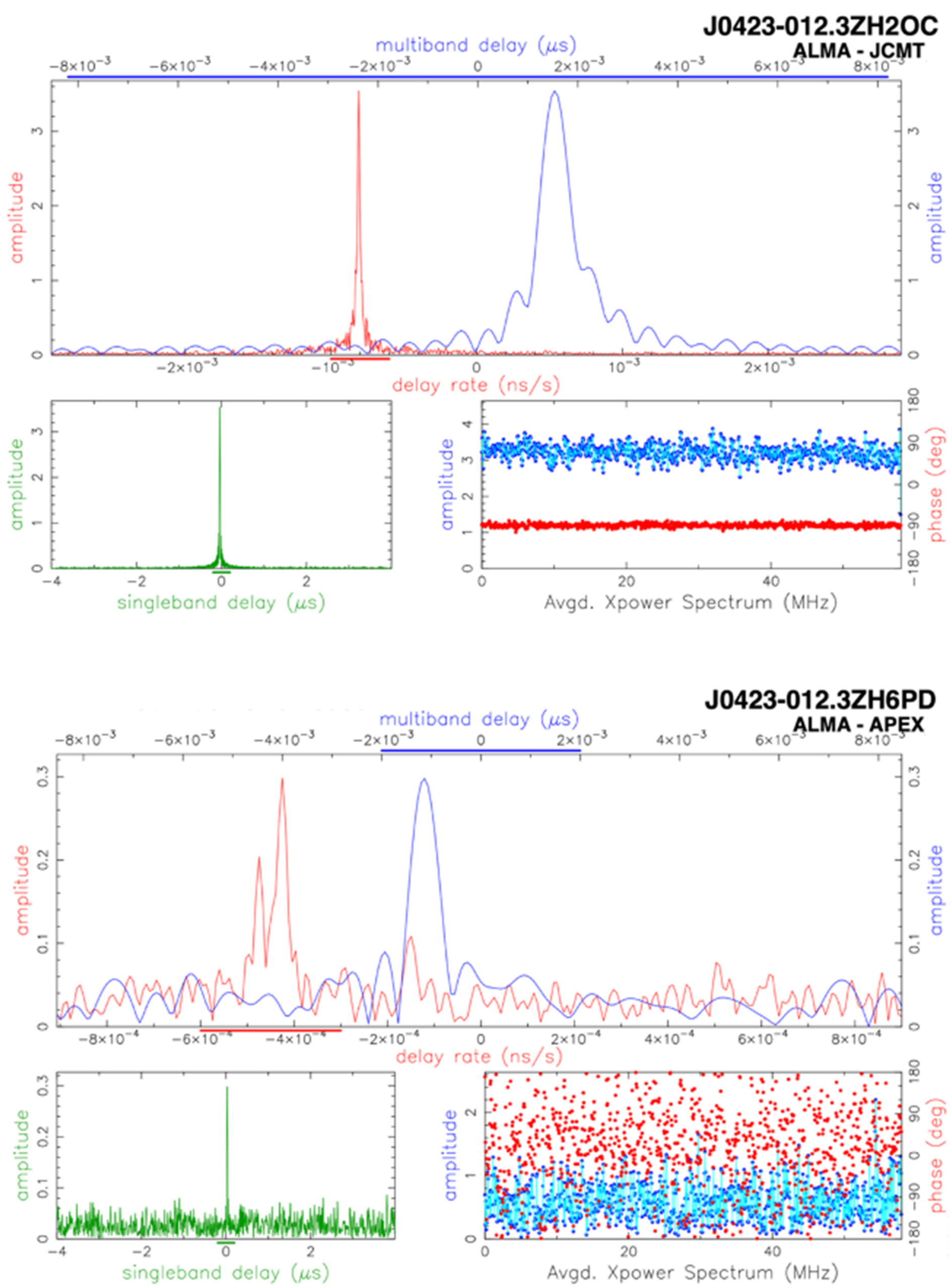

**Figure 5.** *Fourfit* fringe diagnostics illustrating robust and atmosphere-limited detections at 214 and 690 GHz. (Top) ALMA–JCMT baseline at 214 GHz on J0423 −012, showing a strong and unambiguous fringe detection. Each panel consists of three diagnostics: the delay–rate fringe search (top), the single-band delay spectrum (bottom left), and the averaged cross-power spectrum with phase (bottom right). The 214 GHz data exhibit a clear, isolated delay–rate peak, a narrow single-band delay response, and stable cross-power amplitude and phase. (Bottom) ALMA–APEX baseline at 690 GHz on J0423−012, showing a weak but statistically significant fringe detection. An isolated delay–rate peak is present, while the single-band delay response is weak and the cross-power phase is strongly scattered, reflecting rapid atmospheric phase fluctuations and short coherence time at 690 GHz.

In addition, an independent loss is introduced during the formation of the VLBI phased sum. Crew (2020) reported an additional efficiency reduction of approximately 7%, corresponding to a multiplicative factor of 0.93 applied to the VLBI data stream. When these effects are taken into account, a phased array of forty-five ALMA antennas achieves an effective sensitivity equivalent to that of roughly twenty-eight antennas combined ideally.

**Table 3**
Estimated $T_{sys}$ and SEFD at Each Station for Scan-1 and Scan-2

| Estimated System Temperature ($T_{sys}$) and SEFD for Each Station (Scans 1 and 2) | | | | | | |
|---|---|---|---|---|---|---|
| | | | $T_{sys}$ (K) | | SEFD (1000 Jy) | |
| Station | Elev. (deg) | PWV (mm) | 680 GHz | 690 GHz | 680 GHz | 690 GHz |
| ALMA (12 m) Single antenna | ∼45.0 | | 1250 | 1800 | 70.9 | 102 |
| Phased 45 antennas | | 0.46 (avg. ALMA antennas) | | | 2.5 | 3.5 |
| APEX (12 m) | ∼45.8 | 0.46 (same as ALMA) | 950 | 950 | 58.9 | 58.9 |
| JCMT (15 m) | ∼39.3 | 1.3 | 3800 | 4666 | 185 | 228 |

**Note.** The PWV at ALMA and APEX was the Average of the Measurements of the ALMA Antennas. See Text for Details.

**Table 4**
Summary of Correlation Results for the Band 9 VLBI Experiment

| Scan | Frequency (GHz) | Baselines | Baseline Length (M$\lambda$) | Estimated Sensitivity[a] (Jy) | Estimated S/N[a] | Measured S/N |
|---|---|---|---|---|---|---|
| J0423-012 (2.3 Jy) | | | | | | |
| 2 | 685 | AX | 6 | 0.11 | 20 | 12 |
| | | AM | 21,720 | 0.25 | 9 | Non-detect |

**Notes.** The scans list the observed sources, with baselines AX (ALMA–APEX) and AM (ALMA–JCMT). Shown are the baseline lengths and the estimated and measured signal-to-noise ratios (SNRs). The estimated values are the best among the sub-bands, which are centered at 680 GHz (440 $\mu$m). For J0423–0120, a fringe was detected only on the ALMA–APEX baseline in Scan 2, with a measured SNR of 12. No fringes were detected on the ALMA–JCMT baseline.
[a] The SNR is from the 680 GHz subband, with 2 GHz bandwidth and 20 s integration for AX and 10 s for AM.

In the following VLBI sensitivity estimate, we focus on the 680 GHz band, as it provides a more favorable overall likelihood of fringe detection. We consider three primary efficiency factors: the quantization efficiency ($\eta_q$), the coherence efficiency ($\eta_c$), and the polarization efficiency ($\eta_p$)

The standard 2-bit quantization used in astronomical VLBI introduces a signal loss of approximately 12%, corresponding to a quantization efficiency of $\eta_q \approx 0.88$. As discussed in Appendix, the stability of the station frequency standards leads to additional coherence loss. Numerical evaluation of the coherence function at 685 GHz indicates rms coherence values of 0.96 for the ALMA–JCMT baseline and 0.85 for the ALMA–APEX baseline at an integration time of 10 s and 20 s, respectively; these values are adopted as the corresponding coherence efficiencies.

The observations were conducted in linear polarization. Because both JCMT linear feeds were correlated against ALMA, the worst-case loss in correlated amplitude due to an unknown polarization-frame rotation is limited to cos 45° ≈ 0.71. This value is therefore adopted as a conservative lower bound on the polarization efficiency. We verified the polarization labelling by examining the cross-polarization products (XY and YX), which show consistent fringe solutions and no evidence of an X/Y swap. This confirms that the relative polarization mismatch does not exceed 45°, supporting the adopted efficiency limit.

Additional instrumental averaging effects within the correlation chain may also contribute some coherence loss at 690 GHz, although this was not investigated further in the present experiment.

With these efficiency factors taken into account, the VLBI sensitivity between stations A and B can be estimated as follows:

$$\mathrm{SEFD_{VLBI}} = \eta_q \cdot \eta_c \cdot \eta_p \cdot \sqrt{\frac{\mathrm{SEFD}_A \cdot \mathrm{SEFD}_B}{2 \cdot \Delta B \cdot t}}, \tag{2}$$

where $\Delta B$ is the observation bandwidth (2 GHz) and $t$ the integration time. For the ALMA–APEX baseline, we set $t = 20$ s, corresponding to the nominal solution interval adopted above. For the ALMA–JCMT baseline, we set $t = 10$ s, corresponding to the extrapolated reference solution interval. Using the efficiency factors discussed above, the resulting 1$\sigma$ baseline sensitivities are 0.11 Jy for the ALMA–APEX baseline and 0.25 Jy for the ALMA–JCMT baseline.

The primary target, J0423–0120, is an ALMA calibrator with extensive monitoring records. Its flux density was measured at about 2.3 Jy at 674 GHz on both 2024 November 15 and December 14. Incorporating the above efficiency factors, the final expected SNR values for the ALMA–APEX and ALMA–JCMT baselines are summarized in Table 4.

These sensitivity estimates provide a useful reference for interpreting the fringe outcomes. Under the adopted efficiency factors and 20/10 s integration times, the expected baseline sensitivities are broadly consistent with the detection of J0423–0120 on the ALMA–APEX baseline and the non-detection on the ALMA–JCMT baseline.

## 7. Discussion

A central question in evaluating this experiment is how the measured fringe SNRs compared with detection thresholds. In VLBI practice, a SNR ratio of about ∼7 is generally taken as the minimum for robust fringes (TMS). On the ALMA–APEX baseline, the measured SNR was ∼12, above the threshold but 40% below our estimate of 20. This shortfall indicates additional loss terms not captured in the nominal sensitivity estimate. Applying a comparable degradation to the ALMA–JCMT baseline would place its effective SNR below the detection threshold. The non-detection on this baseline is therefore consistent with the sensitivity estimate, although additional baseline-dependent losses on the intercontinental link cannot be excluded.

At 230 GHz, however, the same intercontinental baselines yielded strong fringe detections with high SNR, indicating that the baseline geometry, delay model, and fundamental station performance were sound. Although the available 230 GHz data do not permit a quantitative scaling of sensitivity losses to 690 GHz, these results suggest that the null detection at 690 GHz is unlikely to be caused by gross instrumental or correlator failures. Instead, the result is more naturally explained by frequency-dependent losses, limited high-frequency coherence, or other high-frequency-specific effects affecting the intercontinental link.

Instrumental sensitivity and observing frequency are key factors. Operation at a slightly lower frequency—for example, ∼650 GHz—would offer improved receiver performance and atmospheric transmission, with modest resolution loss. Since SNR scales with both target brightness and sensitivity, such adjustments could significantly enhance detectability.

Target choice is equally important. While J0423–0120 was sufficiently bright, the secondary source was too faint for phasing. A systematic search for brighter, compact calibrators at Band 9 is warranted. Promising candidates from the ALMA catalog include J2253+1608, J2232+1143, J1256–0547, and J1058+0133.

On the other hand, if the non-detection of the primary target on the ALMA–JCMT baseline is due to the source being resolved, one can place a lower limit on the source size by assuming a circular Gaussian brightness distribution. Adopting a conservative detection threshold of $7\sigma$, with $\sigma \approx 0.25$ Jy, on the 22 G$\lambda$ baseline, the correlated flux density of J0423–0120 (total flux density ≈ 2.3 Jy) must fall below ≈ 1.75 Jy. Under these assumptions, the inferred full width at half maximum (FWHM) of the source is required to be $\gtrsim 2.6$ $\mu$as.

Maximizing sensitivity also requires addressing polarization losses. In this experiment, all three stations recorded a single linear polarization. In the worst case, an unknown relative polarization angle can reduce the correlated amplitude by a factor of $\cos 45° \approx 0.71$, corresponding to a conservative ≈29% loss in correlated flux. This loss can be mitigated by converting the data to circular polarization, either using ALMA's PolConvert software (Martí-Vidal et al. 2016) or by installing quarter-wave plates (QWPs) at APEX and JCMT.

QWPs suitable for 600–700 GHz exist in both metamaterial and quartz designs (Arikawa et al. 2008; Tydex Optics[39]). Combining both linear polarizations could in principle improve SNR by $\sqrt{2}$ for unpolarized sources, as demonstrated by the EHT at 230 GHz (EHT Collaboration 2019a), though at 690 GHz the gain will require careful calibration of differential phase and bandpass response.

Atmospheric stability plays a dominant role in determining coherence at 690 GHz. Due to the limited dataset obtained in this experiment, the timescales inferred here should be treated as practical estimates for the present observations. For the detected ALMA–APEX fringe, the data support a practical integration time of about 20 s. For the longer ALMA–JCMT baseline, scaling from the 230 GHz data on the same baseline gives an approximate 690 GHz reference timescale of order 10 s. These values reflect the combined effects of atmospheric stability, ALMA WVR corrections, baseline geometry, and the full VLBI signal chain.

At 690 GHz, rapid tropospheric phase fluctuations are expected to dominate the short-term coherence, but their impact is partly mitigated by the ALMA Phasing System (APS). The ∼16 s phasing cadence sets a practical limit on how well atmospheric variations can be tracked and corrected. Residual phase errors within each interval, together with imperfect phasing efficiency, lead to a progressive loss of coherence on comparable timescales.

In contrast, the 230 GHz data show much weaker coherence loss over the tested integration range (Section 6.1), indicating that the coherence time at that frequency exceeds the available data span. This difference supports the interpretation that the shorter effective coherence time at 690 GHz arises from increased sensitivity to atmospheric phase fluctuations and the finite cadence of the phasing corrections, rather than from instrumental limitations.

Instrumental phase stability, including the frequency standard, contributes a secondary decoherence effect. As discussed in the Appendix, the measured clock stability corresponds to only minor coherence loss over 10–20 s integrations. At 685 GHz, the estimated coherence factors are 0.96 for ALMA–JCMT at 10 s and 0.85 for ALMA–APEX at 20 s, corresponding to amplitude losses of approximately 4% and 15%, respectively. These losses are included as efficiency factors in the sensitivity estimates. Because the corresponding clock-limited coherence times (≈90–150 s) are substantially longer than the 10–20 s timescales adopted or inferred from the VLBI data, clock instability does not set the dominant integration limit at 690 GHz under these observing conditions.

[39] Tydex Optics, THz Waveplates Datasheet, www.tydexoptics.com (accessed 2025).

The limited stability of the APS at 690 GHz is likely tied to this atmospheric regime. During this experiment, APS phasing solutions were produced at a cadence of 16 s, comparable to the 20 s VLBI integration time. When atmospheric phase fluctuations evolve on comparable or shorter timescales, residual intra-integration phase noise can limit coherent averaging. In addition, APS performance degraded on timescales of order 1–2 minutes, comparable to the coherence time (∼100 s) inferred from station frequency-standard stability considerations.

In the sub–terahertz regime, particularly at the highest ALMA frequencies, the tropospheric delay is no longer strictly non-dispersive, with frequency-dependent contributions becoming increasingly relevant near atmospheric molecular resonances (e.g., Curtis et al. 2009; TMS). While WVR-based phase correction effectively mitigates fluctuations associated with the wet component of the atmosphere, its performance can be reduced under very low-PWV conditions where water-vapor fluctuations are small and difficult to measure. In such cases, significant residual phase variations remain—likely associated with dry atmospheric constituents—leading to rapid coherence loss in the phased sum on timescales of order minutes (e.g., Matsushita et al. 2017b).

Consistent with this picture, a short-lived fringe was detected on the ALMA–APEX baseline, but no sustained fringes were observed, and none were detected on ALMA–JCMT despite excellent weather. These results suggest that the APS hardware and software remain fundamentally sound—consistent with proven performance at 230 GHz—but that sustaining coherence at 690 GHz is substantially more demanding. With only a limited number of usable scans, the evidence remains indicative rather than conclusive.

Addressing these limitations will require improved treatment of both dispersive and non-dispersive atmospheric effects at Band 9. Frequency-dependent atmospheric models, such as the Atmospheric Transmission at Microwaves (ATM) code (Pardo et al. 2001b), could be incorporated into the APS pipeline to constrain both dispersive and non-dispersive terms, while faster calibration cycles or interleaved phase referencing could help track rapid fluctuations (e.g., Asaki et al. 2023; Maud et al. 2023). Together with improved delay correction, broader calibrator searches, and dual-polarization recording, such measures could significantly extend APS stability and enhance sensitivity.

This experiment thus marks the first demonstration of VLBI in the 690 GHz atmospheric window. Although no fringes were detected on ALMA–JCMT, the detection on ALMA–APEX, combined with the agreement between measured and predicted sensitivities, validates both the methodology and sensitivity estimates. These results provide confidence that with continued refinements in phasing, calibration, and target selection, near-terahertz VLBI can progress from marginal to routine detections. In this context, modest improvements in coherence retention and effective sensitivity would move single-baseline 690 GHz VLBI into the regime where first-order photon-ring signatures become accessible, as suggested by recent theoretical studies (e.g., Johnson et al. 2020).

Looking ahead, several experimental and technical developments offer clear pathways to accelerate the improvement. Frequency phase transfer from lower-frequency bands offers a promising avenue for extending coherence times at 690 GHz (e.g., Rioja & Dodson 2011), while spectral-line VLBI observations—particularly of bright maser sources—can serve as robust calibration anchors for high-frequency continuum experiments. The eventual inclusion of space-based VLBI elements would further mitigate atmospheric limitations and enable longer baselines at near-terahertz frequencies (e.g., Kardashev et al. 2013). In parallel, the emergence of next-generation large-aperture submillimeter facilities, such as ATLAST-class (e.g., Mroczkowski et al. 2025) or LST-type telescopes, would substantially enhance collecting area and baseline coverage, transforming 690 GHz VLBI from a pathfinder capability into a powerful and routine probe of horizon-scale physics (e.g., Akiyama et al. 2023; Pesce et al. 2024).

## 8. Summary

This experiment represents the first VLBI observations conducted in the 690 GHz atmospheric window. A short-lived fringe detection on the ALMA–APEX baseline confirms that near-terahertz VLBI is technically feasible under excellent weather conditions. The measured SNR (∼12) was consistent with sensitivity expectations after accounting for additional loss terms, validating the instrumental setup, correlation procedures, and baseline geometry.

No sustained fringes were detected on the intercontinental ALMA–JCMT baseline. Comparison with 230 GHz performance demonstrates that this non-detection is unlikely to result from gross instrumental failure, delay-model errors, or fundamental station instability. Instead, the evidence indicates that coherence retention at 690 GHz is substantially more demanding, with additional frequency-dependent loss mechanisms limiting effective sensitivity.

The dominant limitations arise from atmospheric phase stability and phasing-system cadence. The HOPS analysis of the ALMA–APEX detection supports a nominal Band 9 solution interval of order 20 s under the conditions of this experiment, indicating that useful VLBI integrations at 690 GHz are relatively short. Independent clock-stability estimates indicate coherence limits of $\gtrsim$90 s, demonstrating that atmospheric phase fluctuations—rather than maser instability—set the practical integration timescale. With APS solutions generated every 16 s, residual intra-integration phase noise becomes a significant loss term. Under these conditions, even modest degradation (≈30%–40%) relative to nominal sensitivity can reduce a marginal detection to non-detection on long baselines.

Polarization mismatch, single-polarization recording, and possible dispersive atmospheric contributions near molecular resonances further reduce correlated amplitude. Together, these effects roughly account for the reduction of an expected SNR of ∼20 on ALMA–APEX to ∼12 and for the non-detection on ALMA–JCMT.

If the non-detection on ALMA–JCMT reflects source structure rather than purely instrumental loss, the data imply a lower limit of $\gtrsim 4\ \mu$as on the FWHM of J0423–0120 under a circular Gaussian assumption. While not definitive, this constraint illustrates the resolving power already accessible at 22 G$\lambda$.

This experiment demonstrates that 690 GHz VLBI is technically feasible and that the achieved sensitivity is consistent with expectation. The primary limitation arises from atmospheric phase stability and APS cadence rather than from instrumental failure. Modest improvements in coherence retention, polarization handling, and calibration strategy should therefore be sufficient to move future observations from marginal to robust detection.

Future progress will require improved treatment of dispersive atmospheric effects in the APS pipeline, faster phasing updates, dual-polarization capability, and a search of brighter Band 9 calibrators to improve phasing sensitivity. Techniques such as frequency phase transfer from lower bands may extend coherence times by removing the dominant non-dispersive atmospheric component, though residual dispersive effects at Band 9 will still require careful modeling. Even with these incremental improvements, the ultimate coherence limit at 690 GHz remains set by tropospheric phase stability. The eventual inclusion of space-based elements offers additional pathways to extend coherence and baseline length. With these refinements, near-terahertz VLBI can transition from a pathfinder demonstration to a stable platform for horizon-scale imaging and photon-ring studies.

## Acknowledgments

We acknowledge Hung-Yi Pu for his contributions to the development of the GLT science plans that motivated exploration of VLBI observations in the 690 GHz window. We are also deeply grateful to the ALMA management for the observing time allocated to this experiment, conducted under exceptional weather conditions with the participation of 45 ALMA antennas. Given the technical challenges inherent to near-terahertz VLBI, further opportunities to test and refine this emerging capability would be scientifically valuable. The data made available at the ALMA Archive includes a small sample of the recorded VLBI data to allow further investigations of this unique data set. We thank the anonymous referee for careful and constructive comments, which helped us clarify the interpretation of the integration-time dependence and improve the presentation of the coherence analysis.

This work was made possible through the loan of the *Kuntur* 650 GHz (Band 9) receiver provided by the LLAMA collaboration. The name *Kuntur* acknowledges the privilege of installing LLAMA and observing the Universe from the unique natural and cultural site of the Argentine Puna. LLAMA is primarily supported by Secretaría de Innovación, Ciencia y Tecnología (SICyT, Argentina) and the São Paulo Research Foundation (FAPESP; Fundação de Amparo à Pesquisa do Estado de São Paulo, Brazil) through grants 2011/51676-9, 2015/50360-9, 2015/50359-0, and 2021/01183-8. We also acknowledge financial support from the European Union NextGenerationEU RRF M4C2 1.1 project no. 2022YAPMJH.

We thank Martin Boyd, Parth Patel, Paul Carney, Micah Ledbetter, and Abijith Kowligy of Vector Atomic for loaning the iodine clock, Dan Marrone of the University of Arizona for assistance on the loan, Jessica Dempsey of Astron, the Netherlands, for facilitating the transition of the LLAMA Band 9 receiver cartridge. We would also like to thank Kimberly Miskovetz of the EAO, Yu-Nung Su, Sheng-Jun Lin, Cristina Romero-Canizales, and Min Chul Kam of Academia Sinica Institute of Astronomy and Astrophysics (ASIAA; Taiwan), and the staff at the participating observatories, correlation centers, and institutions for their enthusiastic support.

This paper makes use of the following ALMA data:ADS/JAO.ALMA#2011.0.00014.E. ALMA is a partnership of the European Southern Observatory (ESO; Europe, representing its member states), NSF, and National Institutes of Natural Sciences of Japan, together with National Research Council (Canada), Ministry of Science and Technology (MOST; Taiwan), ASIAA, and Korea Astronomy and Space Science Institute (KASI; Republic of Korea), in cooperation with the Republic of Chile. The Joint ALMA Observatory is operated by ESO, Associated Universities, Inc. (AUI)/NRAO, and the National Astronomical Observatory of Japan (NAOJ). The NRAO is a facility of the NSF operated under cooperative agreement by AUI.

APEX is a collaboration between the Max-Planck-Institut für Radioastronomie (Germany), ESO, and the Onsala Space Observatory (Sweden). The SMA is a joint project between the SAO and ASIAA and is funded by the Smithsonian Institution and the Academia Sinica. The JCMT is operated by the East Asian Observatory on behalf of the NAOJ, ASIAA, and KASI, as well as the Ministry of Finance of China, Chinese Academy of Sciences, and the National Key Research and Development Program (No. 2017YFA0402700) of China and Natural Science Foundation of China grant 11873028. Additional funding support for the JCMT is provided by the Science and Technologies Facility Council (UK) and participating universities in the UK and Canada.

The correlator work was partially funded by US NSF awards: AST-2034306 and AST-2535855. Funding for the GLT is partially supported by the Academia Sinica and by the Ministry of Science and Technology, MOST funding codes: 99-2119-M-001-002-MY4, 103-2119-M-001- 010-MY2, and 106-2119-M-001-013, for Taiwan's participation in the ALMA-NA project. The GLT project is also partially supported at the SAO by the Smithsonian Institution. Support for the GLT's Hydrogen Maser frequency standard used for VLBI at the Greenland Telescope was provided through an award to SAO from the Gordon and Betty Moore Foundation (GBMF-5278).

## Appendix
## Evaluation of the Coherence Function

In a practical VLBI experiment, the measured coherence is not determined solely by the stability of the station frequency standards, but rather by the combined effects of atmospheric phase fluctuations, instrumental stability, and the performance of the phasing system. The formalism presented below evaluates the contribution from frequency standards, which provides an upper bound on coherence. In the main text, the empirically measured coherence time reflects the full system behavior, dominated by atmospheric and phasing effects at 690 GHz.

This section summarizes the evaluation of the coherence function following Section 9 of TMS (Rogers and Moran 1981).

$$r = Z(t)\, e^{j\Delta\varphi(t)}, \tag{A1}$$

where $Z$ is the output amplitude and $\Delta\phi$ the phase error from the two VLBI stations. The phase error arises primarily from local atmospheric fluctuations and the frequency standards at each station.

A time-averaged correlator output over an integration time $T$ is given by

$$\bar{r} = \frac{1}{T}\left|\int_0^T Z(t)\, e^{j\Delta\varphi(t)} dt\right|. \tag{A2}$$

Assuming that phase error dominates and the amplitude remains constant, $|Z| = Z0$, we obtain

$$\bar{r} = \frac{Z_0}{T}\left|\int_0^T e^{j\Delta\varphi(t)} dt\right| = Z_0\, C(T), \tag{A3}$$

where $C(T)$ is the coherence function.

The total coherence in a VLBI observation can be expressed conceptually as the product of multiple contributions,

$$C_{\rm tot}(T) \approx C_{\rm atm}(T)\, C_{\rm phase}(T) \quad C_{\rm clock}(T) \tag{A4}$$

**Table A1**
The K Parameters of the Masers and Optical Clock

| | K2 | K0 |
|---|---|---|
| iMaser 167 | 4.5 e-14 | 2.7 e-14 |
| VA Clock | 1.0 e-14 | 3.3 e-14 |
| iMaser 56 | 1.8 e-13 | 3.6 e-14 |
| TMS, 2ed. H(active) | 1.0 e-13 | 3.0 e-14 |

where $C_{\rm atm}$ represents atmospheric phase stability, $C_{\rm phase}$ accounts for residual errors in the phasing system (e.g., finite solution cadence and imperfect corrections), and $C_{\rm clock}$ represents the contribution from station frequency standards evaluated in this appendix.

The mean-square value of $C_{\rm clock}(T)$ can be written as the series expansion of the Allan variance from the clock stability measurements.

$$\langle C^2_{\rm clock}(T)\rangle = \frac{2}{T}\int_0^T\left(1 - \frac{\tau}{T}\right)\exp\{-\pi^2\nu_0^2\tau^2 [\sigma_y^2(\tau) + \sigma_y^2(2\tau) + \sigma_y^2(4\tau) + \ldots]\}\, d\tau \tag{A5}$$

The Allan variance of a typical VLBI frequency standard can be parameterized as

$$\sigma_y^2(\tau) = [K_2^2 + K_1^2 \ln(2\pi f_n)]\tau^{-2} + K_0^2\tau^{-1} + K_{-1}^2 + K_{-2}^2\tau, \tag{A6}$$

where, for the time spans relevant here ($10^{-2}$–$10^2$ s), only the $K_2$ and $K_0$ need to be considered.

We derived the K parameters for the ALMA maser (iMaser 167), the APEX maser (iMaser 56) from the vendor's specifications, and for the Vector Atomic optical clock from our Allan deviation measurements. The numbers are listed in Table A1. For comparison, values from TMS (Table 9.4) are also included. These represent single-station performance. The effective K parameter for a baseline between stations $a$ and $b$ is obtained by combining the individual values in quadrature:

$$K = \sqrt{(K_a^2 + K_b^2)} \tag{A7}$$

Using these parameters, we numerically evaluated the rms coherence function for the ALMA–APEX and ALMA–JCMT baselines at 685 GHz. In the evaluation, we assume that at very short timescales ($\tau < 0.1$ s), the stability is dominated by white noise. Figure A1 presents the results for both baselines, together with a comparison using the TMS parameters. In this context, the coherence times, where $\langle C^2_{\rm clock}(T)\rangle \sim 0.5$, are 92 s and 150 s for ALMA-APEX and ALMA-JCMT baselines, respectively.

For short time scale (<100 s), both ALMA's and JCMT's (VA) clocks have better stability than the APEX's, likely reflecting that APEX's maser is an earlier-model. The rms coherence is 0.96 for the ALMA–JCMT baseline at $\tau = 10$ s and 0.85 for the ALMA–APEX baseline at $\tau = 20$ s. These values were adopted as efficiency factors when estimating overall baseline sensitivities in this report.

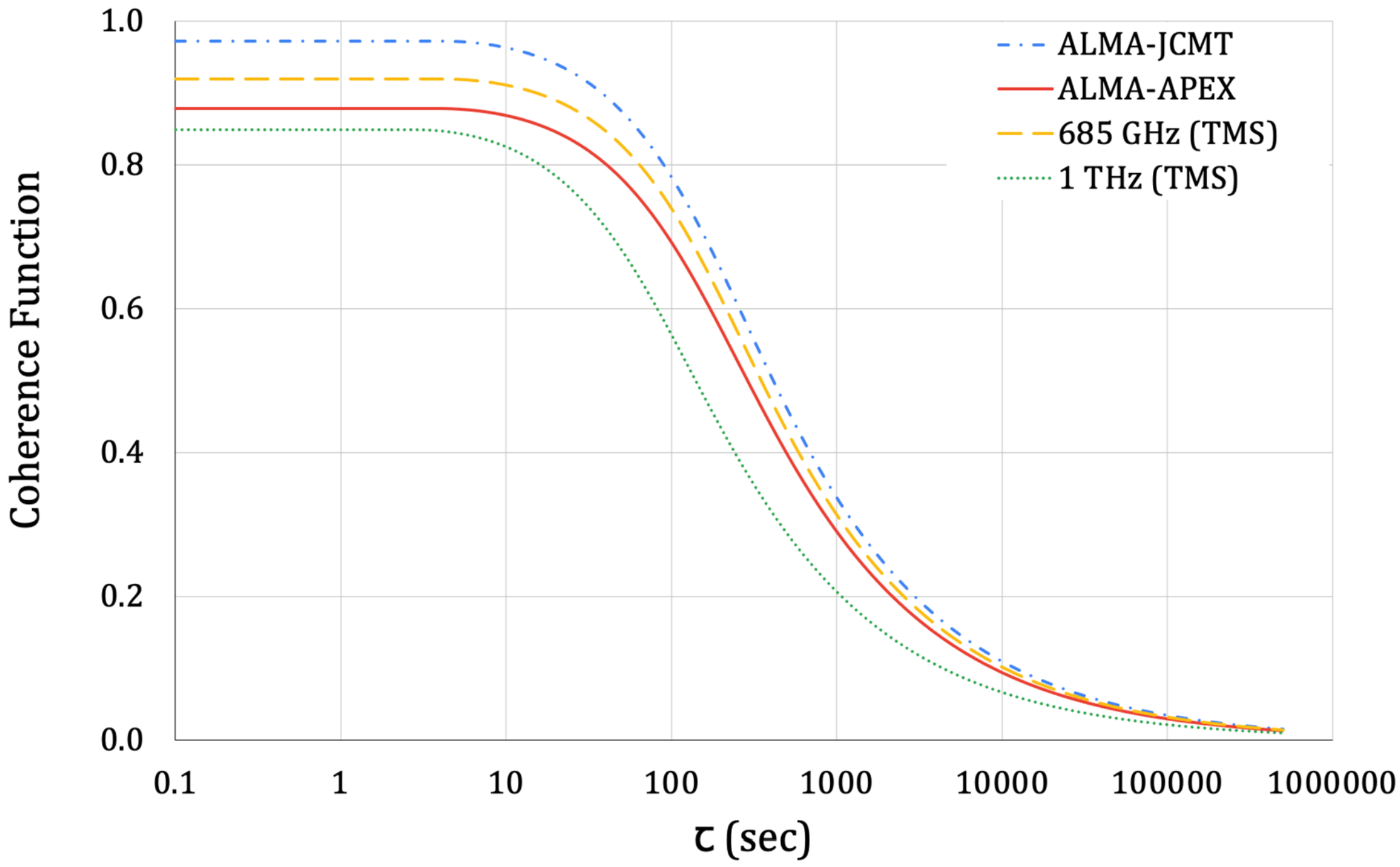

**Figure A1.** Numerical evaluation of the rms coherence function for the ALMA–APEX and ALMA–JCMT baselines at 685 GHz. Results based on measured K parameters (Table A1) are compared with those derived from TMS (Table 9.3). The ALMA–JCMT baseline, using the Vector Atomic optical clock, maintains higher coherence at short timescales (<100 s), with rms values of 0.96 (ALMA–JCMT) at 10 s and 0.85 (ALMA–APEX) at 20 s.

These values indicate that the coherence loss due to frequency standard stability is modest, corresponding to amplitude reductions of approximately 4% (ALMA–JCMT) and 15% (ALMA–APEX). The corresponding clock-limited coherence times of order 100 s are substantially longer than the empirically inferred coherence time of ∼20 s from the VLBI data. This demonstrates that frequency standard stability does not set the dominant coherence limit in this experiment.

At 690 GHz, the dominant limitation on coherence is expected to arise from rapid tropospheric phase fluctuations, together with the finite response of the ALMA Phasing System. The APS applies phase corrections at a cadence of approximately 16 s, which limits its ability to track faster atmospheric variations. Residual phase errors within each interval lead to progressive decorrelation on comparable timescales. There are a number of options available within the existing ALMA Phasing System that could be better tuned in view of what we have found in this investigation. In particular, a shorter phasing cadence and shorter-duration target scans interleaved with atmospheric calibrations would have produced better performance. The 10–20 s operational timescales inferred from the data are consistent with a regime in which atmospheric fluctuations and phasing cadence, rather than frequency-standard stability, dominate the coherence loss.

## ORCID iDs

Ming-Tang Chen https://orcid.org/0000-0001-6573-3318
Geoffrey B. Crew https://orcid.org/0000-0002-2079-3189
Satoki Matsushita https://orcid.org/0000-0002-2127-7880
Geoffrey C. Bower https://orcid.org/0000-0003-4056-9982
Ciriaco Goddi https://orcid.org/0000-0002-2542-7743